\documentclass[letter]{aa}
\pdfoutput=1
\usepackage[utf8]{inputenc}
\usepackage{float}
\usepackage{epsf,graphics,graphicx}
\usepackage[varg]{txfonts}
\def\lsim{\!\!\!\phantom{\le}\smash{\buildrel{}\over
 {\lower2.5dd\hbox{$\buildrel{\lower2dd\hbox{$\displaystyle<$}}\over
                                 \sim$}}}\,\,}
\def\gsim{\!\!\!\phantom{\ge}\smash{\buildrel{}\over
{\lower2.5dd\hbox{$\buildrel{\lower2dd\hbox{$\displaystyle>$}}\over
                               \sim$}}}\,\,} 
\def\asec{\ifmmode ^{\prime\prime}\else$^{\prime\prime}$\fi}

\begin{document}

\title{Linear radio size evolution of $\mu$Jy populations}
\author{M. Bondi\inst{1} 
        \and G. Zamorani\inst{2}
        \and P. Ciliegi\inst{2} 
        \and V. Smol{\v c}i{\'c}\inst{3}
        \and E. Schinnerer\inst{4}
        \and I. Delvecchio\inst{3}
        \and E.\,F. Jim\'enez-Andrade\inst{5}\inst{6}
        \and Daizhong Liu\inst{4}
        \and P. Lang\inst{4}
        \and B. Magnelli\inst{5}
        \and E.J. Murphy\inst{7}
        \and E. Vardoulaki\inst{5} }

\institute{INAF - Istituto di Radioastronomia, Via Gobetti 101, 40129, Bologna
  \and INAF - Osservatorio di Astrofisica e Scienza dello Spazio di Bologna, Via Gobetti 93/3, 40129, Bologna
  \and Department of Physics, Faculty of Science, University of Zagreb, Bijeni{\v c}ka cesta 32, 10002, Zagreb, Croatia
  \and Max-Planck-Institut f\"ur Astronomie, K\"onigstuhl 17, 69117 Heidelberg, Germany
  \and Argelander Institut f\"ur Astronomie, Universit\"at Bonn, Auf dem H\"ugel 71,
  D-53121 Bonn, Germany
  \and International Max Planck Research School of Astronomy and Astrophysics at the
Universities of Bonn and Cologne, Bonn, Germany
  \and National Radio Astronomy Observatory, 520 Edgemont Road, Charlottesville,
   VA 22903, USA
}  

\abstract{We investigate the linear radio size properties of the $\mu$Jy populations
  of radio-selected active galactic nuclei (AGN) and star-forming galaxies (SFGs) using a 
  multi-resolution catalog based on the original VLA-COSMOS 3\,GHz 
 0\farcs75 resolution  mosaic and its convolved images (up to a resolution of 2\farcs2). The final catalog contains
   6\,399 radio sources above a 3\,GHz total flux density of $S_T>20$ $\mu$Jy (median $<S_T>=37$ $\mu$Jy),
   with redshift information (median $<z>=1.0$), and multi-wavelength classification as SFGs, radio-excess AGN
   (RX-AGN), or non-radio-excess AGN (NRX-AGN). RX-AGN are those whose radio emission
exceeds the star formation rate derived by fitting the global spectral energy distribution. 
 We derive the evolution with redshift and luminosity of the median linear sizes of each class of objects.
We find that RX-AGN are compact, with median sizes of $\sim$ 1-2 kpc and increasing with redshift,
corresponding to an almost constant angular size of 0\farcs25.
NRX-AGN typically have radio sizes a factor of 2 larger than the RX-AGN.
The median radio size of SFGs is about 5 kpc up to $z\sim 0.7$, and it decreases beyond this redshift.
Using luminosity-complete subsamples of objects, we separately investigate the effect of redshift and luminosity
dependance.
We compare the radio sizes of SFGs with those derived in the rest-frame far-infrared (FIR) and UV bands.
We find that SFGs have comparable sizes (within 15\%) in the radio and rest-frame FIR, while the sizes measured
in the UV-band are systematically larger than the radio sizes.
}
\maketitle

\section{Introduction}

While the {\it Hubble Space Telescope} (HST) has revealed the morphological properties and sizes in the
UV and optical rest-frame for galaxies up to redshift $\sim 7$
\citep[e.g.,][]{Ferguson04, Hathi08, Oesch10, Ono13, 2014ApJ...788...28V, 2015ApJS..219...15S},
the radio size distribution in the submillijanksy ($\mu$Jy) regime  for different galaxy types is currently not  well known.
This information is crucial in several aspects.
The determination of the linear radio size evolution as a function of redshift in star-forming galaxies (SFGs)
traces the spatial extent of star formation through cosmic time; this is a key ingredient for understanding 
galaxy formation and evolution and, for instance, for testing the paradigm of inside-out growth \citep[e.g.,][]
{Mo1998}. Moreover, given that one problem affecting the measurement of the UV/optical sizes of SFGs is
the obscuration of the inner regions \citep[e.g.,][]{2016ApJ...817L...9N}, radio sizes provide complementary
measurements that are not affected by obscuration.
For active galactic nuclei (AGN), the size of the radio-emitting region  provides
information on the radio emission mechanisms in different types of AGN and is possibly related to the
accretion efficiency of the central black hole. Distinguishing the contributions of the AGN
and SF-related radio emission is critical for investigating the feedback mechanisms. 

Furthermore, an accurate knowledge of the intrinsic radio size distribution is necessary to correct
the observed differential source counts and luminosity functions for incompleteness due to the resolution bias
\citep[e.g.,][]{Bondi03}.  
Differential source counts derived from different surveys show a scatter that  is larger than the quoted errors
for flux densities $\lsim 200$ $\mu$Jy \citep[e.g.,][]{dezotti2010, Padovani2016}, which, in addition to cosmic
variance, is generally explained as being due to the different corrections applied
to compensate for the resolution bias in different surveys \citep{Condon2007, Heywood2013}. 
We can expect that SFGs and AGN have different radio size distributions and consequently different
corrections for resolution bias, even in the same flux density range.
This is also relevant for the predictions of the numbers of SFGs and AGN for future surveys with the Square Kilometer Array (SKA)
and its precursors \citep[e.g.,][]{Novak18, Mancuso17, Bonaldi18}. 

The determination of radio sizes of $\mu$Jy sources is challenging, and previous studies have suffered
from small statistics, incomplete classification that did not separate SFGs from AGN, and a correlation
between the estimated sizes and angular resolution.
\begin{figure*}[htp]
 \centering
 \includegraphics[width=7.0cm]{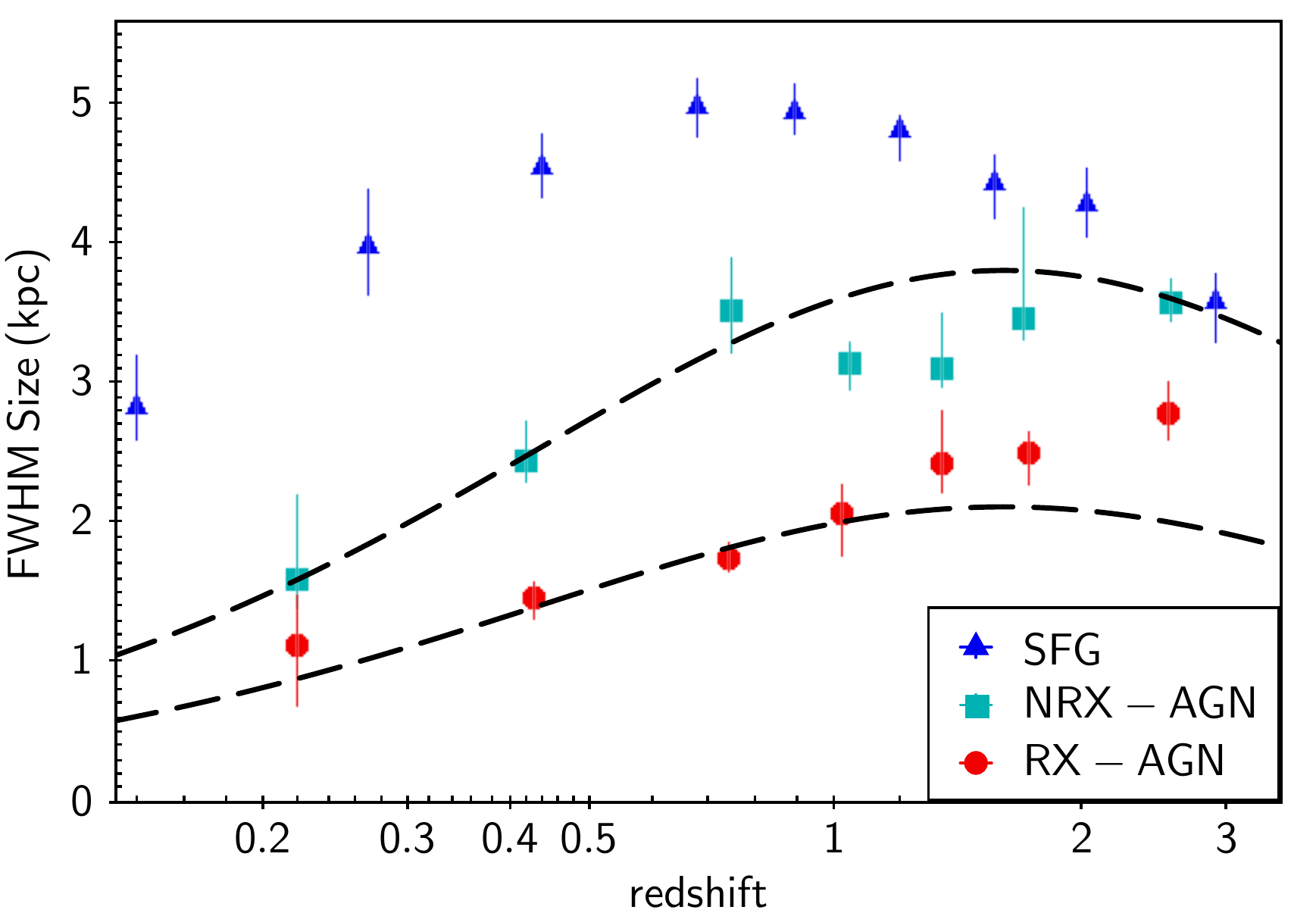}
 \includegraphics[width=7.0cm]{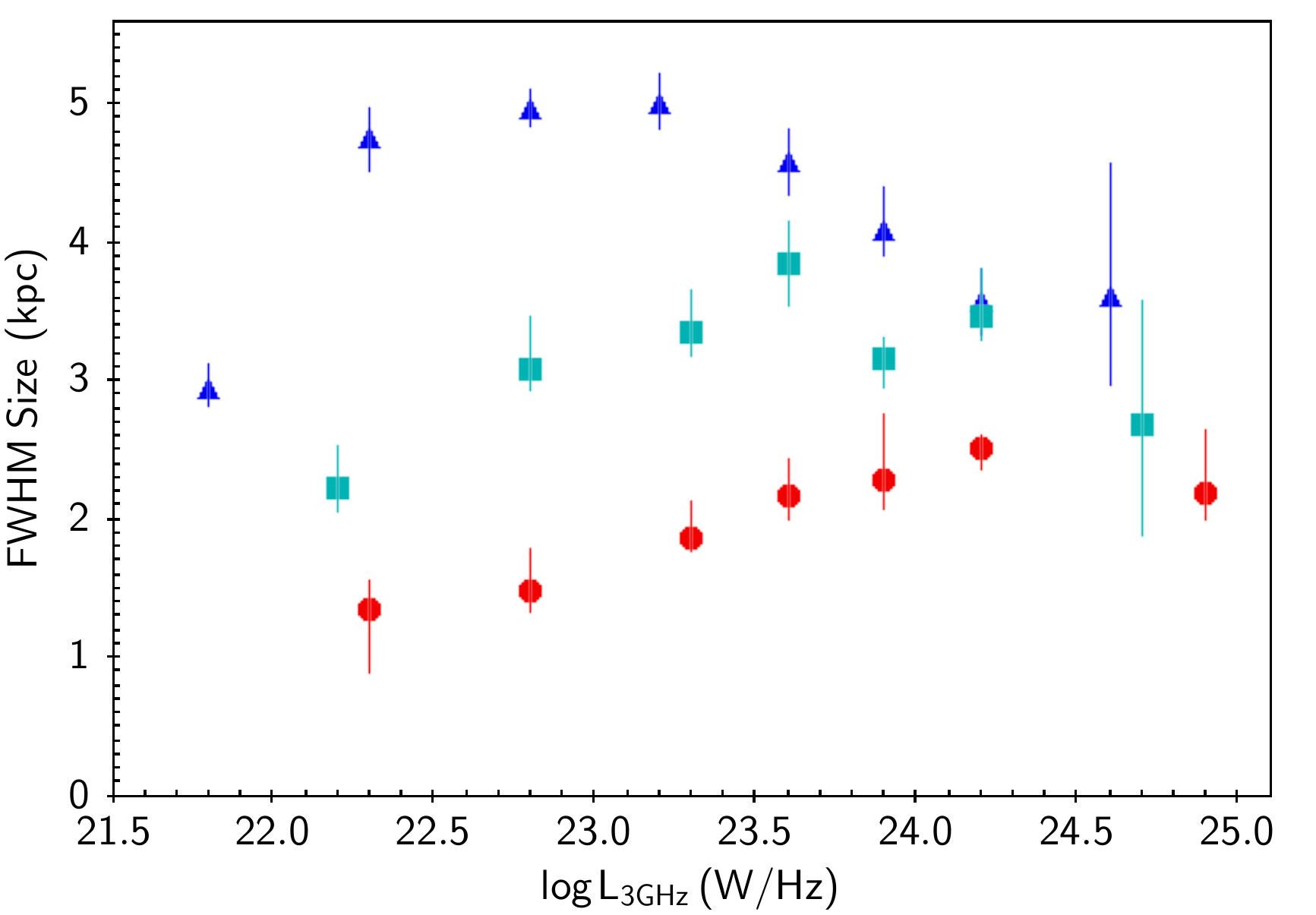}
\caption{Median linear size as a function of redshift (left panel) and 
3\, GHz radio luminosity (right panel) for the different classes of $\mu$Jy radio sources
(see the inset legend and text for details). The dashed lines in the left panel correspond to a constant angular size 
of 0\farcs 25 (lower line) and 0\farcs45 (upper line).}
 \label{fig:size}
\end{figure*}
\begin{figure*}[htp]
 \centering
 \includegraphics[width=7cm]{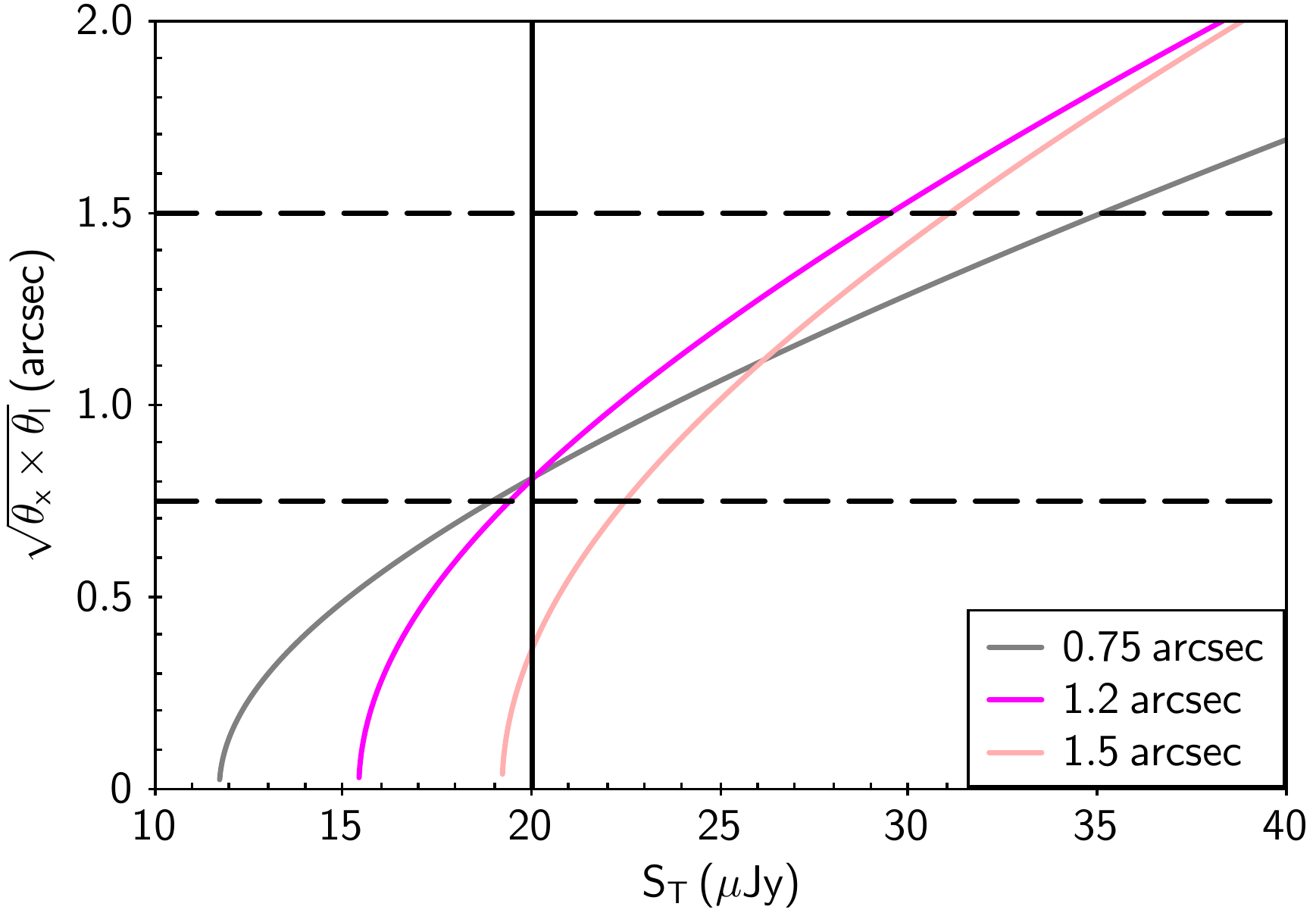}
 \includegraphics[width=7cm]{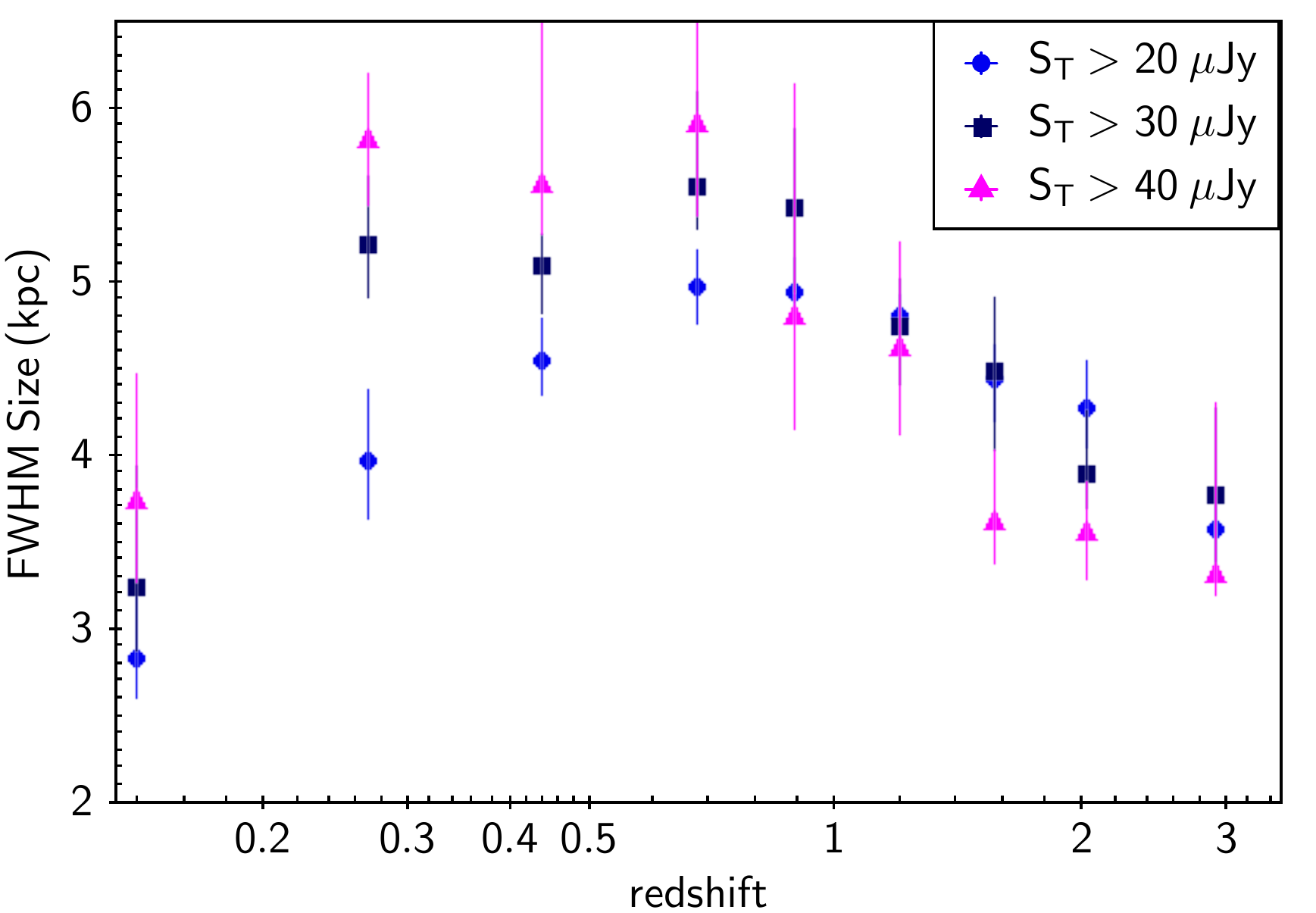}
 \caption{Left panel: Maximum detectable angular size resulting from the combined effects of r.m.s sensitivity 
   and total flux density selection limit for images with 0\farcs75 (grey line), 1\farcs2 (magenta line) and 1\farcs5 resolution
   (pink line).
The maximum angular size is estimated from the ratio $S_T/5\sigma$ using the geometric mean of the 
two limiting cases (see text). Right panel:
Redshift evolution of the median linear sizes of three different samples of SFGs: the blue points 
 are the same as in Fig.\ref{fig:size} and correspond to radio sources with 3\,GHz total flux density $S_T>20$ $\mu$Jy, while
 the purple squares and magenta triangles correspond to subsamples with $S_T>30$ $\mu$Jy and 
$S_T>40$ $\mu$Jy, respectively.
 }
 \label{fig:theta}
\end{figure*}
For instance, \citet{2017ApJ...839...35M} derived a median FWHM major axis of $\sim 0\farcs17\pm 0\farcs03$ from
a sample of 32 sources imaged with the Very Large Array (VLA) at 10\,GHz with a resolution of $\simeq$ 0\farcs22 and a sensitivity
of 0.57 $\mu$Jy\,beam$^{-1}$.
When the same image tapered to 1\farcs0 resolution is used, the median size increases to $0\farcs75\pm 0\farcs17$
(see Table 2 in Murphy et al. 2017). The larger sizes of the sources derived from the tapered image can be
due to the lower resolution, but for some sources, it is likely that faint extended emission is resolved out in the
full resolution, while it is detected in the tapered image, as suggested by the comparison of the total flux
densities at both resolutions.
In the same field, \citet{2017MNRAS.471..210G} used sensitive ($\sigma\simeq 3$ $\mu$Jy\,beam$^{-1}$)
5.5\,GHz observations with a resolution of $\sim$ 0\farcs5 and obtained a catalog of 94 radio sources
classified as AGN or SFGs using X-ray luminosity and IR colors, finding a median FWHM major axis of
0\farcs2-0\farcs3 for the AGN and 0\farcs8 for the SFGs.
Recently, \citet{Cotton18} derived the angular size distribution of  $\mu$Jy radio sources from a sample
of $\sim 800$ sources obtained from sensitive ($\sigma\sim 1$ $\mu$Jy\,beam$^{-1}$)
VLA observations at 3\,GHz of the Lockman Hole with $3\farcs0$ and $0\farcs66$ resolutions.
The angular sizes are estimated from the brightness ratio at the
two resolutions, and should be regarded as equivalent to circular Gaussian FWHM.
\citet{Cotton18} find that the median size is $\sim 0\farcs3-0\farcs4$ (see Table 4 in their paper).

This Letter presents for the first time the linear radio size properties and its redshift and luminosity evolution
for a large sample (more than 6\,000) of $\mu$Jy radio sources classified as AGN or SFGs.

\section{A multi-resolution catalog of radio-selected SFGs and AGN}

The 3\,GHz multi-resolution (MR) catalog is compiled using the full-resolution (0\farcs75) catalog and
catalogs obtained from radio images convolved to coarser resolutions. The MR catalog
lists 6\,399 radio sources with flux densities above 20 $\mu$Jy, a spectroscopic or photometric redshift,
and a  multi-wavelength classification as SFGs\footnote{In this paper we use SFG for the objects
classified as clean SFG in \citet{2017A&A...602A...2S}, i.e., sources without a radio excess.}
or AGN \citep{2017A&A...602A...3D, 2017A&A...602A...2S}.
AGN are further divided into radio-excess (RX) or non-radio-excess (NRX) sources.
RX-AGN are those whose radio emission
exceeds (by more than $3\sigma$ at any given redshift) the star formation rate (SFR) derived by fitting
the global spectral energy distribution \citep[][hereafter D17]{2017A&A...602A...4D}.
The total number of sources in each class is listed in Table\,\ref{tab_2}.
The details on how the MR catalog is obtained are given in Appendix\,\ref{A1}, and the procedure
used to derive the angular size of the radio sources is described in Appendix\,\ref{A2}.

\begin{figure*}[htp]
 \centering 
 \includegraphics[width=7cm]{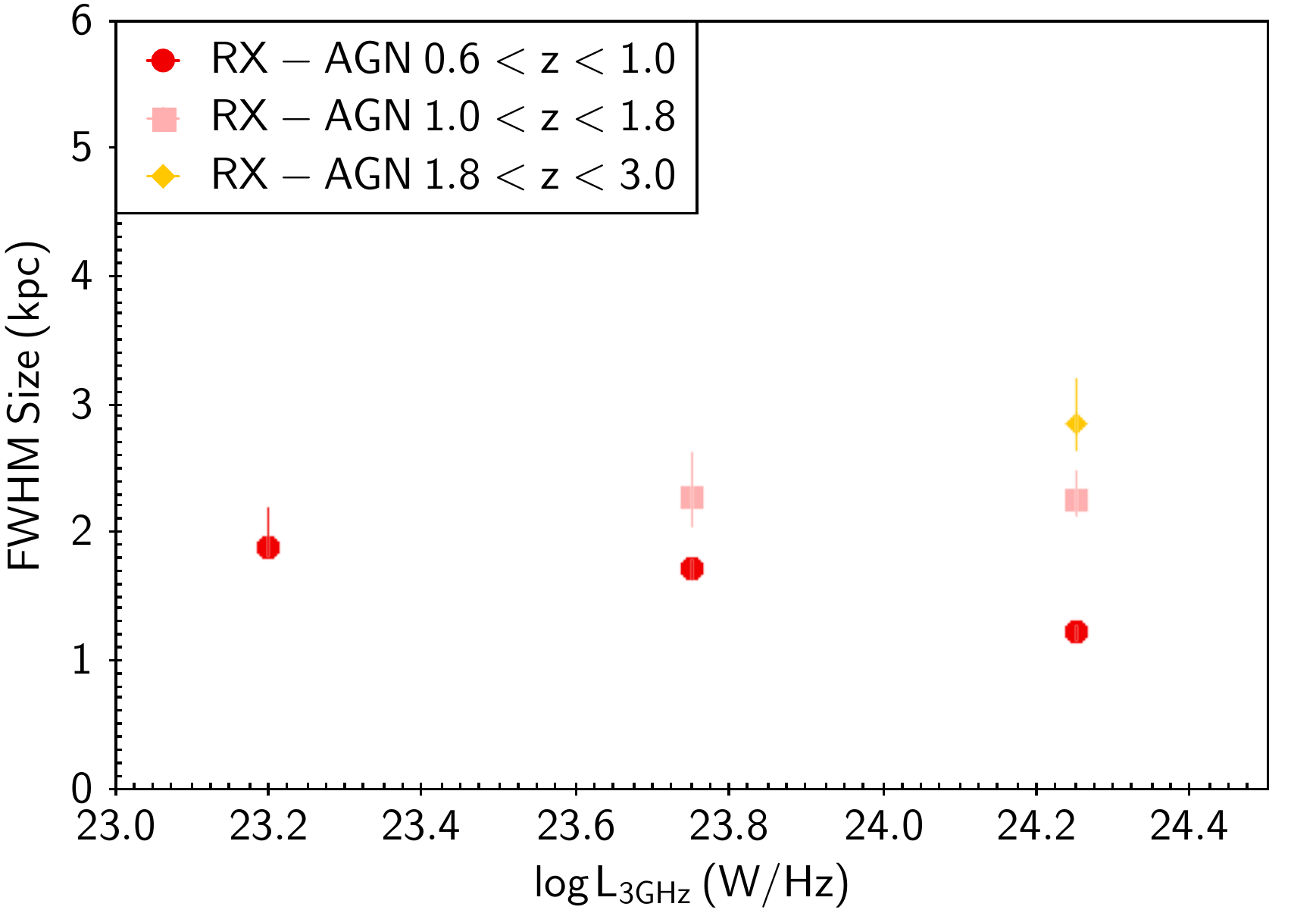}
 \includegraphics[width=7cm]{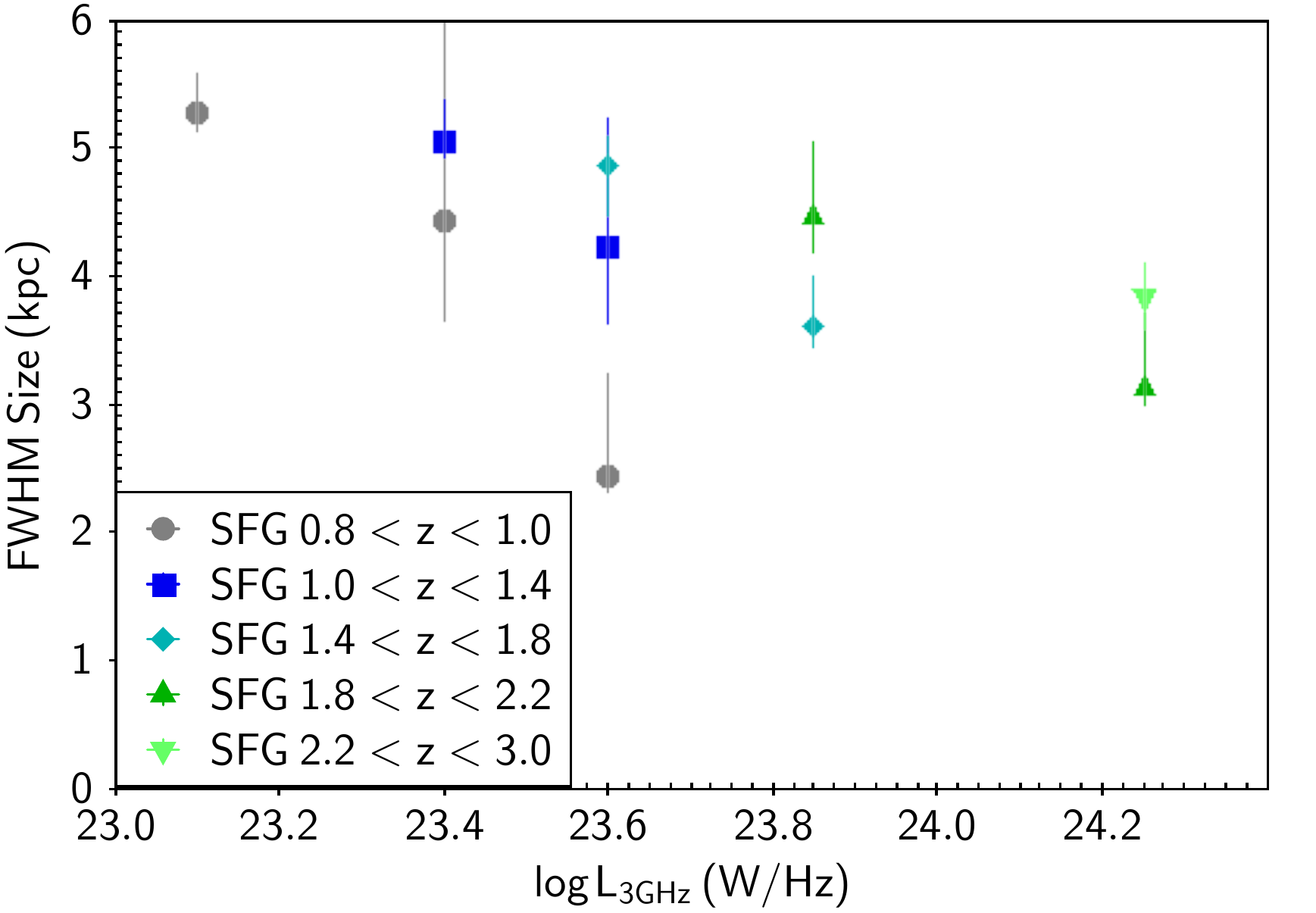}
 \caption{Median FWHM radio size for a subsample of RX sources (left panel) and SFGs (right panel)
   complete in 3\,GHz luminosity (L$_{\rm 3GHz}$) in the redshift bins listed in the legends.}
 \label{fig:rx_size_complete}
\end{figure*}

\section{Linear size evolution of $\mu$Jy populations}

Sources in each class (SFG, NRX-AGN, and RX-AGN) were binned in redshift and 3\,GHz total luminosity
(L$_{\rm 3GHz}$), and for each bin, the Kaplan-Meier median estimator of the FWHM linear size was
derived using  the {\tt ASURV} statistical package that implements the methods described by
\citet{1985ApJ...293..192F} and \citet{1986ApJ...306..490I} to properly handle censored data.
The median sizes with the  $1\sigma$ errors  are reported in Tables\,\ref{tab-a1}\,and \ref{tab-a2} and
are shown in Fig.\,\ref{fig:size}.

The RX- and NRX-AGN have significantly different sizes of up to $z\sim 2.5$ or $L_{\rm 3GHz}\sim 10^{24}$ WHz$^{-1}$.  
RX-AGN are systematically more compact with median FWHM values, increasing with redshift or
luminosity, in the range $\sim$ 1-2 kpc corresponding to an almost constant angular size of 0\farcs 25
(see the lower dashed line in the left panel of Fig.\ref{fig:size}).
The median linear size of NRX-AGN steadily increases from $z\sim 0$ up to $z\sim 0.7$ 
(or to $L_{\rm 3GHz}\sim 10^{23}$ WHz$^{-1}$), converging to a roughly constant value of $\sim$ 3-4 kpc
up to the highest redshift (luminosity).  
NRX-AGN are a factor of $\sim 2$ more extended than RX-AGN up to $z\sim 1$ or $L_{3GHz}\sim 10^{23.5}$,
and the median size is consistent with a constant angular size of 0\farcs45 (see the upper dashed line
in Fig.\,\ref{fig:size}).

The median linear size of the SFGs has a peak at $z\sim 0.6$ (or $L_{\rm 3GHz}\sim 10^{22.5}$ W\,Hz$^{-1}$),
corresponding to a value of $\sim$4.8 kpc, followed by a steady decrease at higher redshift or luminosity.
The rise of the median linear size of SFGs up to $z\sim 0.6$ is, at least partly, the result of the resolution bias.
Even if our  MR catalog corrects partly for this effect, this approach is limited by the lower sensitivity of
the convolved images with respect to the full resolution image (see Table 2 in D17). 
The solid lines in Fig.\ref{fig:theta} (left panel) show the expected maximum angular size for a $\ge 5\sigma$
detection as a function of total flux for three convolved images.
This size is calculated using the geometric mean of Eqs. (2) and (3) in
\citet{2017A&A...602A...1S}. The plot shows that when only sources with $S_{\rm T}\ge 20$ $\mu$Jy
(as for our MR catalog) are selected, we can recover  all the sources with an intrinsic size  of $\lsim$ 0\farcs8.
To detect sources as large as 1\farcs5 (corresponding to a linear size of 4.8 kpc at $z\sim 0.2$), we
could use the 1\farcs2 resolution image, but we would need to raise the completeness limit to $S_{\rm T}\gsim 30$ $\mu$Jy.

Therefore, even using the convolved images, we can expect to miss sources with intrinsic sizes  of
$\sim 5$ kpc at low ($z<0.5$) redshift.
We tested this conclusion by restricting our sample to sources with $S_{\rm T}>30$ $\mu$Jy  and
$S_{\rm T} > 40$ $\mu$Jy and repeated the analysis for the SFGs and NRX-AGN.
The results for the SFGs are shown in Fig.\ref{fig:theta} (right panel): for $z>1$ the median sizes are not
significantly affected by different cuts in  $S_{\rm T}$, while for $z\lsim 0.7$ we indeed find
larger median sizes using only sources with  $S_{\rm T}>30$ $\mu$Jy. A threshold of $S_{\rm T}>40$ $\mu$Jy
does not produce  a further significant change in the median sizes.
The NRX-AGN exhibit no significant differences in the median sizes regardless of the cut in the total
flux density. 
Therefore, while we find evidence of an increase in the median radio linear sizes for the NRX-AGN
from $z\sim 0$ to $z\sim 0.7,$ the radio size of the $\mu$Jy population of SFGs is consistent with being
constant at a value of $\sim$ 5-6 kpc up to $z\sim 1$ followed by a steady decrease thereafter.

These results confirm, with a much improved statistics, the sizes derived from \citet{2017MNRAS.471..210G}
for SFGs and AGN.
If we consider all our sources together (no distinction between SFGs and AGN), as done in \citet{Cotton18}
and \citet{2017ApJ...839...35M}, we obtain a median linear size of $3.58\pm 0.05$ kpc that corresponds
to $\simeq$ 0\farcs45 at $z=1$ (the median redshift of our sources). This value is comparable with that
found by \citet{Cotton18} when we take into account that we measure the size along the major axis, while they
derive an equivalent circular size.
The more compact sizes found by \citet{2017ApJ...839...35M} can be partly intrinsically due to the energy-dependent propagation of electrons producing the synchrotron emission at 10\,GHz and 3\,GHz
\citep[e.g.,][]{Murphy2012} and partly due to selection effects caused by the higher resolution (0\farcs22)
of the 10\,GHz observations.

\section{Discussion}

\subsection{AGN sizes}
We find that the radio emission of the $\mu$Jy RX-AGN is dominated by a compact radio source ($\sim$ 0\farcs25),
most likely the radio nucleus. 
This is consistent with the results obtained from Very Long Baseline Array (VLBA) observations
\citep{2017A&A...607A.132H}.
The MR catalog includes 405 RX-AGN detected by the 1.4\,GHz VLBA observations at milliarcsec resolution
\citep{2017A&A...607A.132H}.
While the percentage of RX-AGN is only 26\% in the entire MR catalog, almost 87\% of the sources
detected by the VLBA observations are RX-AGN. The VLBA observations are biased toward the most compact sources.
Both RX- and NRX-AGN detected by the VLBA have median sizes of $\sim 1.5$ kpc.
The VLBA-detected RX-AGN have the same median size as the entire RX-AGN population in the MR catalog.
Conversely, the NRX-AGN with a VLBA detection are about a factor of 2 more compact than the NRX-AGN
in the MR catalog, and  as noted by \citet{2017A&A...607A.132H}, they display a systematically higher
radio emission than that expected from star formation, but not enough to be above the
threshold we assumed to be classified as RX-AGN.

\begin{figure}[htp]
 \centering
 \includegraphics[width=7cm]{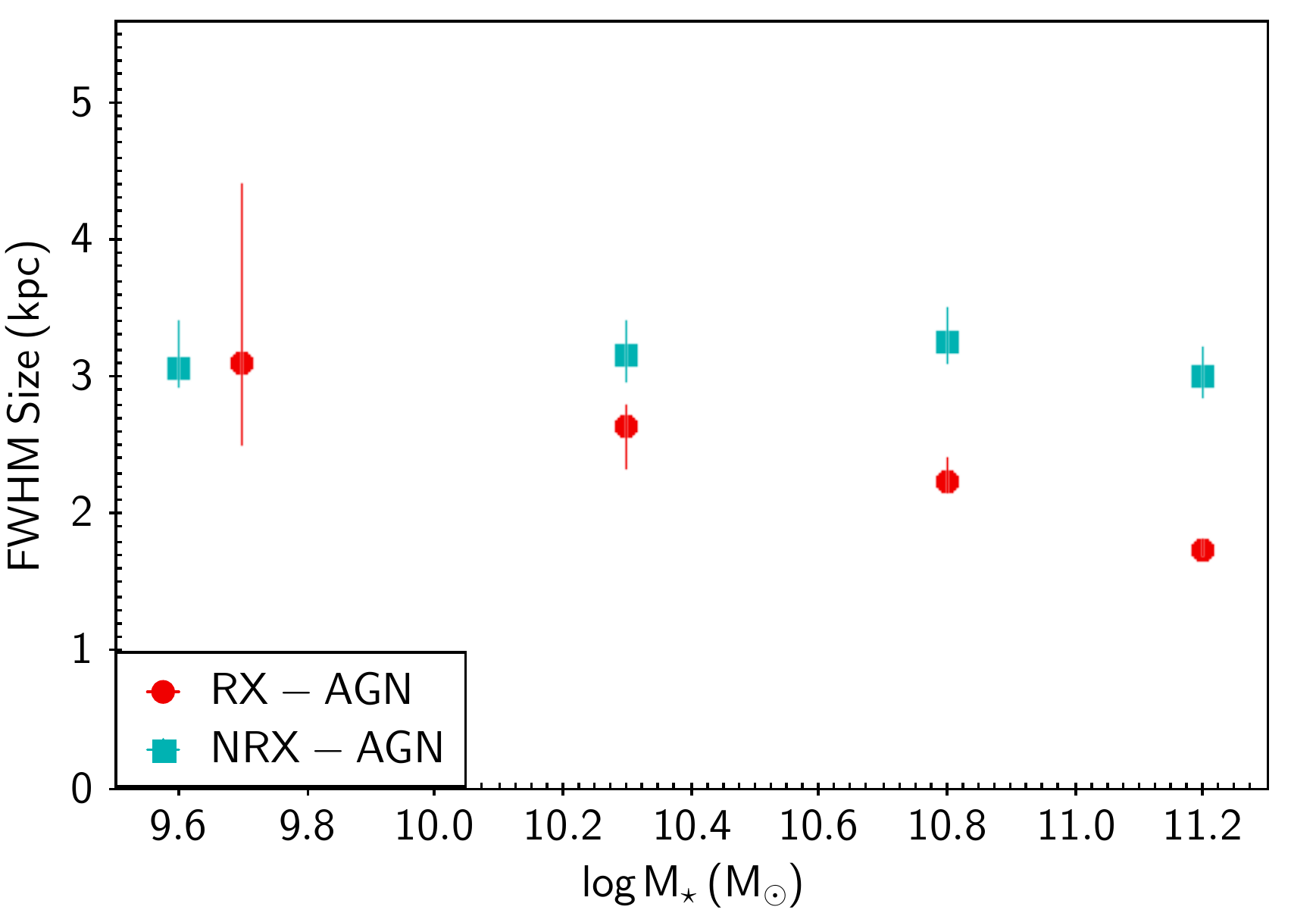}
 \caption{Median FWHM radio size of AGN with RX-AGN and NRX-AGN as a function of the host galaxy stellar mass.}
 \label{fig:mass}
\end{figure}

To separate the effects of L$_{\rm 3GHz}$ and redshift on the median radio sizes, 
we selected subsamples of RX-AGN fully complete in L$_{\rm 3GHz}$ at three different redshift intervals
($0.6<z<1.0$, $1.0<z<1.8$, $1.8<z<3.0$). The complete 
bins in L$_{\rm 3GHz}$ are $10^{22.9}<L_{\rm 3GHz}<10^{23,5}$, $10^{23.5}<L_{\rm 3GHz}<10^{24.0}$,
$10^{24.0}<L_{\rm 3GHz}<10^{24.5}$. 

The median linear radio size derived from the luminosity-complete subsamples of RX-AGN are shown in 
Fig.\,\ref{fig:rx_size_complete} (left panel). We find that redshift is the dominant parameter ruling the
increase in radio sizes.
At lower redshifts ($z\lsim 1$), RX-AGN are compact ($<2$ kpc), and the most luminous sources are the most compact ones.
These properties are consistent with those of FR0 sources observed
in the local Universe  \citep{Baldi15}. FR0 sources are expected to be more numerous at $z\sim 1$ \citep{Whittam16}
and in surveys selected at frequencies above 1.4 GHz. We find that $\sim 25\%$ of the $\mu$Jy sources in our
MR catalog can indeed be regarded as FR0, a significant percentage of the $\mu$Jy population that should be considered
when modeling the $\mu$Jy radio sky \citep{Whittam17}. Moreover,
the median sizes of subsamples that are complete at higher luminosity confirm the overall trend of Fig.\,\ref{fig:size}:
RX-AGN at $z\gsim 2$ have significantly larger radio sizes ($2.9^{+0.35}_{-0.20}$ kpc for $10^{24.0}<L_{\rm 3GHz}<10^{24.5}$),
  comparable to those of NRX-AGN, than RX-AGN at $z\lsim 1$ with the same L$_{\rm 3GHz}$. 
This is consistent with the bulk of RX-AGN at $z\gsim 2$ being radiatively efficient AGN (see Appendix\,\ref{mr3}) hosted in
blue SFGs, as recently found by \citet{Delvecchio18}, in contrast to what is observed at $z\lsim 1,$ where RX-AGN are instead
radiatively inefficient  AGN within red and passive galaxies.
The more extended radio emission in $z\gsim 2$ RX-AGN can be produced by star formation and/or by a
radio jet component that might be able to develop only in objects with higher redshift or luminosity.
The two processes are not mutually exclusive and might be related through AGN feedback \citep[e.g.,][]{Silk13}.
It is interesting to note that the median radio size of $z\gsim 2$ RX-AGN is comparable to the optical size of the
quiescent galaxies at $z\sim 2$ \citep{szomoru12, vandokkum15}. 
Furthermore,  Fig.\,\ref{fig:mass} confirms that the most compact RX-AGN are hosted in the most massive galaxies
\citep{Sadler89, WH91, Nyland16, Baldi18b}, while median linear radio sizes of  RX-AGN are larger in galaxies with lower
stellar mass, suggesting a possible link between the build-up of a massive
galaxy and the presence of a compact radio core.

The accessible range in radio luminosity of NRX-AGN here is too limited to repeat the same analysis on luminosity-complete
subsamples as performed on the RX-AGN.
The NRX-AGN do not show a decrease in radio size in the most massive objects (Fig.\,\ref{fig:mass}),
suggesting that the build-up of the galaxy is not associated with the emerging of a compact and bright
radio core. The median radio size in NRX-AGN is consistent with radio emission (or at least most of it)
being spread across the galactic disk and probably not being related to the AGN detected from the
multi-wavelength diagnostics, but rather originating by star formation processes.

\subsection{SFGs sizes}

Figure\,\ref{fig:size} shows a significant decrease in the median radio size  with redshift (or luminosity)
in SFGs for $z\gsim 1$.
To separate the dependence on redshift and luminosity, we considered subsamples of sources complete in
$L_{\rm 3GHz}$ at a given redshift, as was done for the RX-AGN.
The results are shown in Fig.\,\ref{fig:rx_size_complete} (right panel). We note that at a fixed redshift, the median
radio size decreases with increasing $L_{\rm 3GHz}$: SFGs with higher $L_{\rm 3GHz}$ have a more centrally peaked surface
brightness that can be either due to a nuclear starburst or a low-luminosity radio AGN that is not detected by the
multi-wavelength diagnostics. This is in agreement with the higher fraction of AGN found in the most extreme
  starburst galaxies \citep[e.g.,][]{Bonzini15}.
Figure\,\ref{fig:rx_size_complete} also shows a hint for more extended SFGs at higher redshifts
at a given  $L_{\rm 3GHz}$.

To compare the radio sizes of SFGs to those derived in the rest-fame FIR and optical-UV bands by
\citet[][F17]{2017ApJ...850...83F}, \citet[][T17]{2017ApJ...834..135T}, and  \citet[][S15]{2015ApJS..219...15S},
we converted the measured FWHM of the radio major axis into the circularized effective radius (R$_e$, see Appendix\,\ref{A3}). 

F17 and T17 derived the rest-frame far-infrared (FIR) size of the star-forming regions that are obscured by large amount of dust
using ALMA observations of 1034 and 12 sources, respectively. The SFGs from F17 are split
into three redshift bins ($z=$1-2, 2-4, and 4-6), while the 12 sources from T17 are all within $2.2<z<2.5$.

In the optical-UV, the most updated and comprehensive study of galaxy R$_e$ is presented by S15
using  HST observations for a sample of $\sim 190\,000$
galaxies at $z=0-10$. This study is an extension of that published by \citet{2014ApJ...788...28V}, who
provided similar and consistent results.
S15 derived a best-fit relation to the median size evolution with redshift for 
three different ranges of UV luminosity for different statistics (median, average, and mode).
We used the median fit obtained for the UV-bright galaxies ($-24 < M_{\rm UV}< -21$), corresponding
to stellar masses in the range $10.0 \lsim \log(M/M_\odot)<11.0$. This is  
consistent with the median stellar mass of $\log(M/M_\odot)=10.6$ derived for the hosts of our radio-selected
SFGs.

\begin{figure}[htp]
 \centering
 \includegraphics[width=7cm]{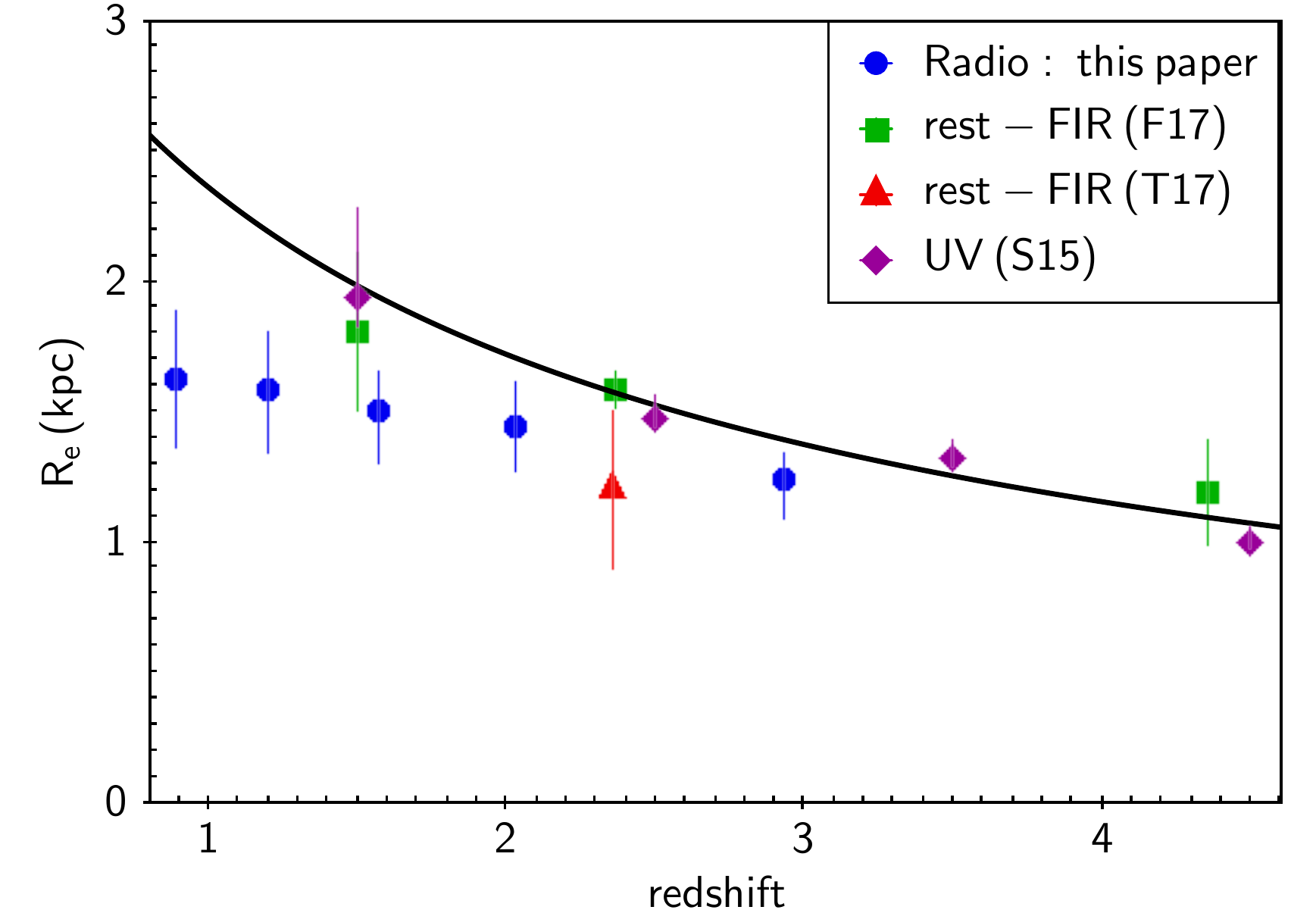}
 \caption{Redshift evolution of the circularized effective radius R$_e$ for SFGs with $z>1$. Blue filled circles are the radio
   R$_e$  from our sample of radio-emitting SFGs, magenta triangles and green  squares are the rest-frame FIR 
   R$_e$ from \citet[][F17]{2017ApJ...850...83F} and \citet[][T17]{2017ApJ...834..135T}, purple diamonds are the UV median
   values from \citet[][S15]{2015ApJS..219...15S}. The errors on the radio R$_e$ also include the dispersion of the axial ratio
   median value we used to convert into circularized effective radius.   
The solid line is the best-fit median size evolution derived for R$_e$ in the UV-band for UV-luminous SFGs 
by S15, given by R$_e=B_z(1+z)^{\beta_z}$, with $B_z=4.05$ and
$\beta_z=-0.78$.
}
 \label{fig:size_comparison}
\end{figure}

In  Fig.\,\ref{fig:size_comparison} we plot the median $R_e$ values derived in the radio, FIR, and UV bands,
and the best-fit median size evolution curve obtained by S15 ($R_e=B_z(1+z)^{\beta_z}$, with $B_z=4.05$ and $\beta_z=-0.78$).

The radio sizes derived from our sample of SFGs are systematically smaller (by less than $15$\%), but agree
within the errors with the sizes measured in the rest-frame FIR by F17.
We  note that for spiral galaxies in the local Universe the radio sizes are generally larger than those measured
  in the FIR \citep[e.g.,][]{Bicay1990, Murphy2006, Murphy2008}, because the synchrotron-emitting electrons are diffused out of the galactic disk, but at high redshift, the inverse-Compton losses will dominate
the diffusion, and the radio emission is expected to be confined within the star-forming regions.
  A similar difference is observed to the rest-frame UV sizes for SFGs at $z\gsim 2$, while the offset between
  the radio and UV sizes increases at lower redshifts, becoming 30\% at $z\sim1$.
UV measurements may be affected by dust obscuration, in particular, an increased dust attenuation in the central
regions as has been suggested to be typical in SFGs at $z\sim 2$ \citep{Tacchella18},
would produce larger UV sizes than those that could be recovered without a significant radially dependent obscuration.

\section{Summary}
We have presented an analysis of the linear radio size evolution for a sample of more than 6\,000 $\mu$Jy radio sources
classified as SFGs, non-radio-excess AGN (NRX-AGN) and radio-excess AGN (RX-AGN) from the COSMOS
3\,GHz Large Program Survey, using a multi-resolution radio catalog. The results can be summarized as follows:
\begin{itemize}
\item  We find compelling evidence that RX-AGN, NRX-AGN, and SFGs
 at a $\mu$Jy flux density level have distinct median linear sizes with
 different redshift (or luminosity) evolutions. The sizes of RX-AGN increase with redshift (or 3\,GHz radio luminosity)
 up to $z\sim 2$, with values in the range 1-2 kpc, corresponding to an almost constant angular sizes of
 0\farcs25. NRX-AGN show systematically larger sizes than those of RX-AGN, with a more complex evolution with
 redshift or luminosity. From $z\sim 0$ to $z\sim 0.6,$ the median linear size of RX-AGN steadily
 increases up to $\sim$ 3-4 kpc and is consistent with being constant at higher redshifts. 
 SFGs, after correcting for the resolution bias, have a larger and approximately constant median size ($\sim 5$ kpc)
 from $z\sim 0.2$ up to $z\sim 1$, followed
 by a decrease at higher redshifts. At $z\sim 2.5,$ the median size of SFGs is comparable to that of the NRX-AGN.
 \item 
 Our results are consistent, but with much improved statistics, with those recently
   reported by \citet{2017MNRAS.471..210G} and \citet{Cotton18}, while we find larger sizes than those derived at 10\,GHz
   with 0\farcs2 resolution by \citet{2017ApJ...839...35M}.
   This difference is probably partly intrinsic because of the energy-dependent propagation of synchrotron
-emitting electrons at 10\,GHz and 3\,GHz, and is partly due to selection effects caused by the higher angular resolution
   of the 10\,GHz observations.
\item
  To distinguish the effect of redshift and luminosity, we have selected subsamples of sources 
  that are complete in $L_{\rm 3GHz}$ at each redshift bin. Considering only RX-AGN in the redshift range $0.6<z<1.0$,
  we find that higher luminosity objects are more compact. Moreover,  at a given $L_{\rm 3GHz}$ , the RX-AGN at higher
  redshift are more extended than those at lower redshift. The median radio size of 
  RX-AGN at $z\sim 2.5$ is a factor of 2.4 larger than the median size of local (z$\lsim 1$) RX-AGN. While the
  radio emission in local RX-AGN originates from a compact radio core, at high redshift ($z\gsim2$), the RX-AGN
  are significantly more extended in the radio, and part of this radio emission could be associated with star formation.
  The median radio size of high-redshift RX-AGN ($\simeq 2.9$ kpc) is comparable to the optical size of quiescent
  galaxies at $z\sim 2$. We also find a clear anticorrelation between the stellar mass content and the radio sizes in
  RX-AGN: smaller radio sizes are  present in the most massive hosts, while the less massive (and typically high
-redshift) galaxies have larger radio sizes. The radio sizes of NRX-AGN do not significantly depend on the stellar
    mass of the host galaxy.
\item
  Considering only SFGs with $z>0.8$ that are complete in $L_{\rm 3GHz}$ at each redshift, we find that  the most luminous SFGs
  are more compact, while at a given  $L_{\rm 3GHz}$ , the SFGs at higher redshift are more extended.
 
 \item
 The radio linear sizes of SFGs at $z\gsim1$ agree (within 15\%) with the rest-frame FIR sizes
 published by F17 and T17. The same is true for the rest-frame UV sizes of SFGs at $z\gsim 2$, while at lower redshifts
 the discrepancy is larger (at $z\lsim 1$ radio sizes are $\sim 30\%$ smaller than UV sizes).

 \end{itemize}

\begin{acknowledgements}
VS acknowledges support from the European Union's Seventh Frame-work program under grant
agreement 337595 (ERC Starting Grant, "CoSMass”).
BM and EJA acknowledge support by the Collaborative Research Centre 956,
sub-project A1, funded by the Deutsche Forschungsgemeinschaft (DFG).
EJA acknowledges support of the Collaborative Research Center 956, subproject A,
funded by the Deutsche Forschungsgemeinschaft (DFG).
\end{acknowledgements}  
\bibliographystyle{aa} 
\bibliography{paper-size.bib} 

\begin{appendix}
\section{Building a multi-resolution catalog}
\label{A1}
The VLA-COSMOS 3\,GHz Large Project produced a final image with a median  {\it rms} sensitivity
of 2.3 $\mu$Jy\,beam$^{-1}$ over a field of 2 square degrees with a resolution of 0\farcs75  
and a catalog of about 11\,000 radio sources $\geq 5\sigma$ \citep{2017A&A...602A...1S}.
This catalog allowed us to derive the radio source counts down to a flux density of about 10 $\mu$Jy
\citep{2017A&A...602A...1S,2017A&A...602A...2S} and even deeper, extrapolating from the derived luminosity
function for different classes of objects \citep{2017A&A...602A...6S,Novak18}.
The subarcsecond resolution of the VLA-COSMOS 3\,GHz observations is an advantage for a proper
optical counterpart identification, but has two relevant disadvantages affecting radio sources
with intrinsic angular sizes $>$0\farcs75:

\begin{enumerate}
\item For moderately resolved sources, the total flux density (and consequently the angular
  size) in the catalog can be underestimated.
\item   A significant number of heavily resolved sources can be completely missed
  because the catalog is selected on the peak flux density.
\end{enumerate}

The original full-resolution mosaic has been convolved, producing a set of images and
associated catalogs, with increasingly coarser resolutions. All the catalogs were generated
using the software {\tt blobcat} developed by \citet{2012MNRAS.425..979H}, as detailed in
\citet{2017A&A...602A...1S}.

To derive an estimate of the size of the radio-emitting region, we decided to adopt the following procedure.
The first step was to compile a multi-resolution (MR) catalog based on the full-resolution
catalog and on the catalogs derived from the convolved images to recover the correct flux and size of
the moderately resolved sources (see Appendix\,\ref{mr1}), and as many as possible of the heavily
resolved ones that  are undetected in the full-resolution 3\,GHz catalog (Appendix\,\ref{mr2}).
The program {\tt blobcat} does not provide a measure for the size, therefore the second step was to fit all the
resolved sources in the MR catalog, assuming a Gaussian brightness distribution, using the
prescription that the peak and total fluxes from the fits must be consistent with the values derived
by {\tt blobcat} (Appendix\,\ref{A2}).

\subsection{Sources detected at full resolution}
\label{mr1}
We started from the 3 GHz radio source counterpart catalog containing all the 9\,161 objects
detected with a signal-to-noise ratio (S/N) greater than 5 at a resolution of
0\farcs 75  and an optical/infrared (OIR) counterpart assigned either in the COSMOS2015 \citep{2016ApJS..224...24L}, 
{\it i}-band \citep{Capak07}, or IRAC catalogs \citep{Sanders07}, as described in \citet{2017A&A...602A...2S}.
We selected 8\,864 sources within the region of 2 square degrees with right ascension (RA) and declination (DEC) in the range 
$149.41134<$ RA (deg) $<150.8271$ and  $1.49862<$ DEC (deg)$<2.91286$.
In this region, 59 radio sources are classified as multiple, and they were treated differently,
leaving a total of 8\,805 single-component sources.

For each of these sources we identified counterparts (best match within a 1\arcsec\, search radius) 
in the $5\sigma$ radio catalogs obtained from the convolved radio mosaics at resolutions of
0\farcs 90, 1\farcs 20, 1\farcs 50, 1\farcs 80, and 2\farcs 20.
When a source had a counterpart in a lower resolution catalog (starting from 0\farcs 90), we
calculated the difference between the total flux densities  reported in the two catalogs (lower $-$ higher resolution).
If this difference was $> 2\times \sigma_{\rm comb} = 2\times\sqrt{\sigma_{\rm LR}^2 + \sigma_{\rm HR}^2}$ 
, where $\sigma_{\rm comb}$ is defined through $\sigma_{\rm LR}$ and $\sigma_{\rm HR}$,
the total flux density errors of the lower and higher resolution catalogs,  then the lower resolution was selected as
the best resolution. For each source, we iterated this comparison to all the counterparts in catalogs with increasingly
lower resolutions (from 0\farcs90 to 2\farcs20). In this way, we associated a best resolution and a corresponding
total flux density with each source detected at full resolution.
This is the MR catalog for the single-component radio sources.
About $20\%$ of the sources detected at full resolution have a best resolution $>$0\farcs 75. 
The number of sources whose total flux density is affected by this procedure and the best
resolution adopted in the MR catalog is reported in Table\,\ref{tab_1}.
  The same procedure could not be applied to the 59 multi-components sources.
For these sources the total flux in the MR catalog was measured on the 1\farcs5 resolution convolved image using the same method
as we employed to derive the flux for the original catalog. 

\begin{table}
\caption{Number of sources at each resolution in the multi-resolution catalog}
\label{tab_1}
\centering
\begin{tabular}{lllllll}
\hline\hline
 Best res. & 0\farcs 75 & 0\farcs 90 & 1\farcs 20 & 1\farcs 50 & 1\farcs 80 & 2\farcs 20 \\
\hline
\#         & 7,155     &  968      & 365       & 170       & 85        & 62       \\
\hline
\end{tabular}
\end{table}

\begin{figure*}[htp]
 \centering
 \includegraphics[width=6.5cm]{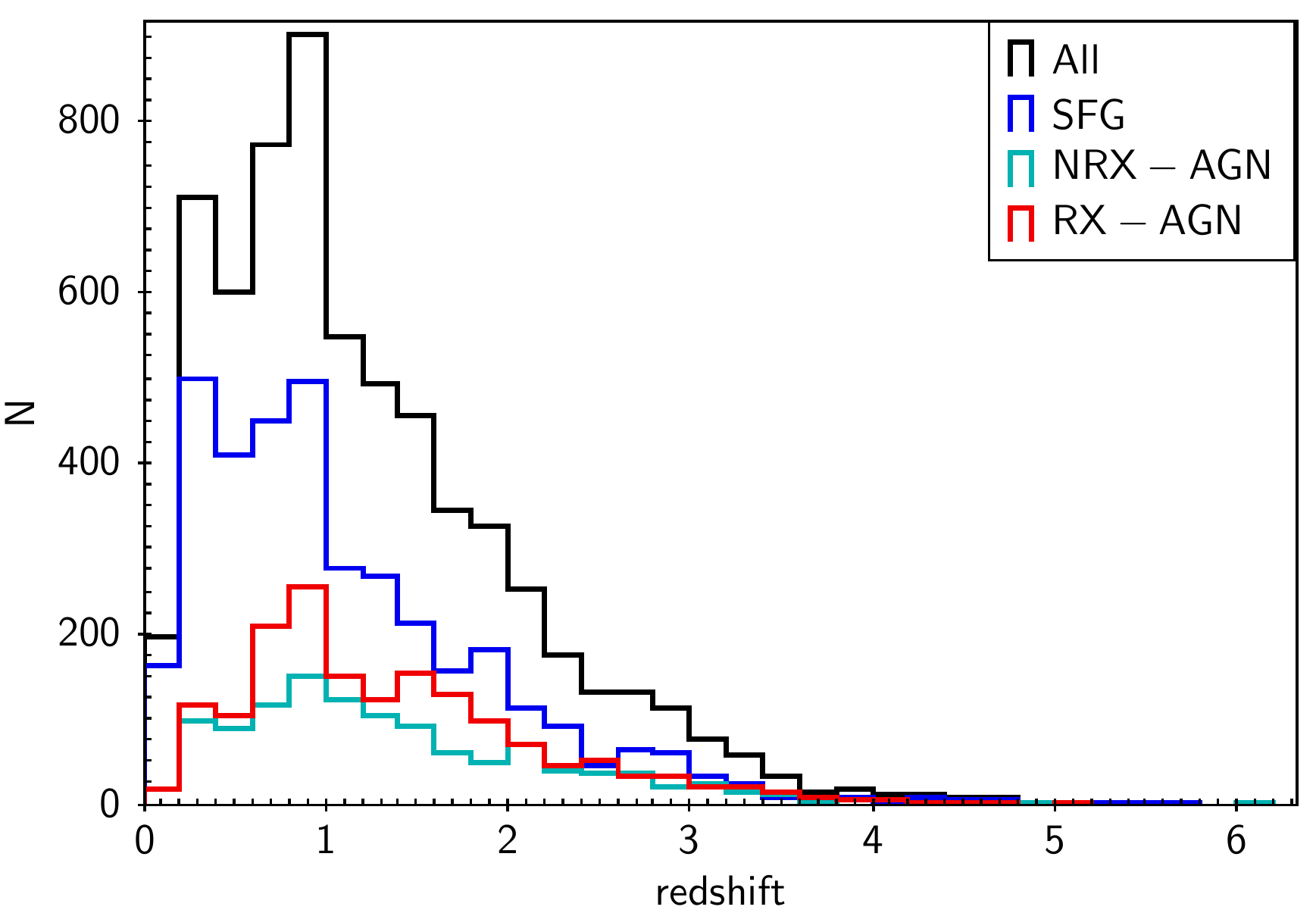}
\includegraphics[width=6.5cm]{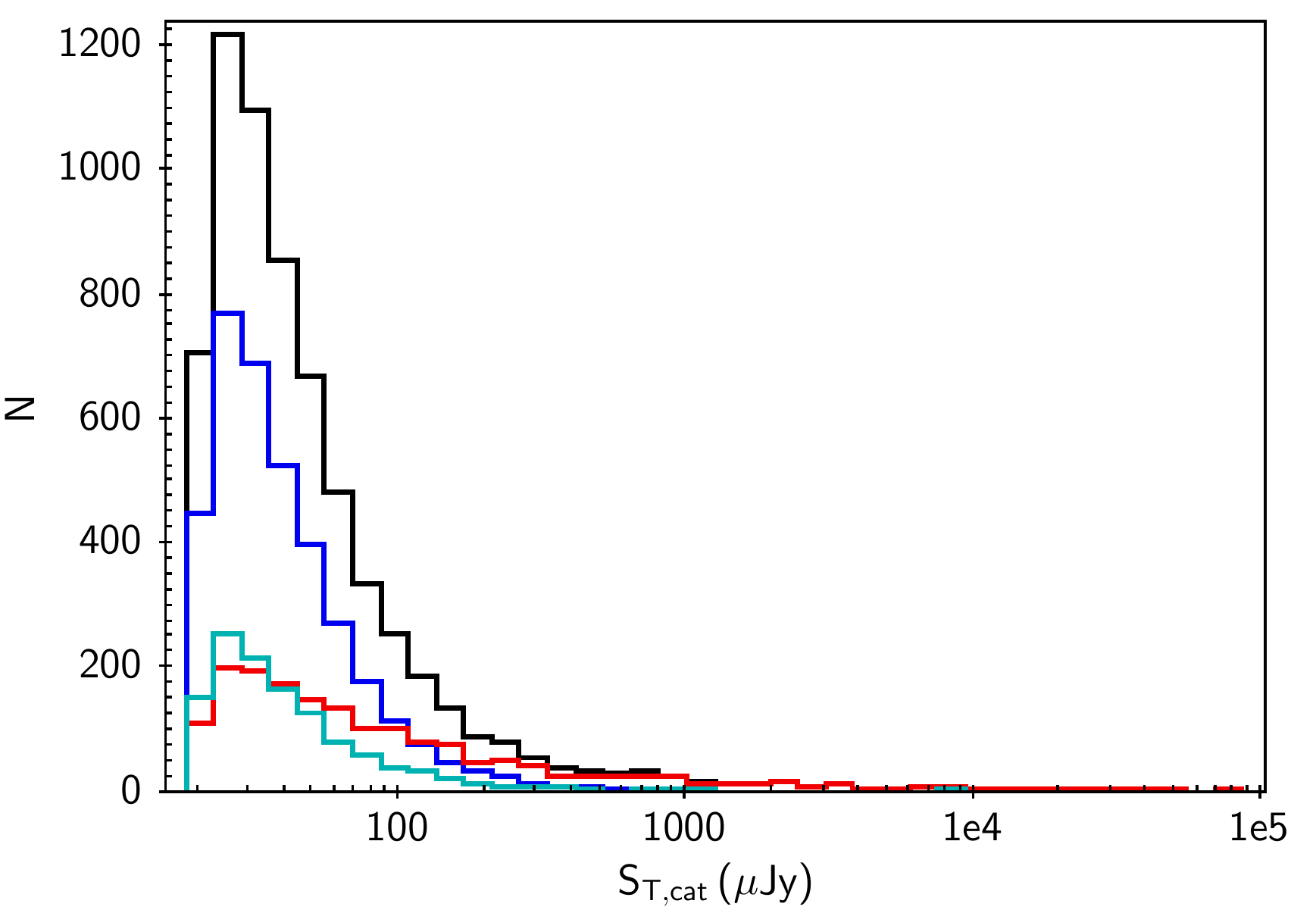}
\caption{Redshift and 3\,GHz total flux density distributions for 6\,399 radio sources in the final multi-resolution
catalog.
The black line shows the distribution of the whole sample, while color-coded lines show the distributions
of different populations as reported in the inset legend.}
 \label{fig:dist1}
\end{figure*}

\subsection{Sources undetected at full resolution}
\label{mr2}
So far, we have only considered the radio sources with OIR counterparts
detected in the full-resolution mosaic. As previously noted and discussed in D17,
a non-negligible number of sources with intrinsic sizes $ >$0\farcs 75 could be missing from
this catalog as a result of the resolution bias. As a first step, we followed the method used by D17
selecting 449 infrared selected sources with a $>5\sigma$
radio counterpart in at least one of the catalogs of the convolved radio images up to a
resolution of 2\farcs 2. These IR sources are  from a 24 $\mu$m prior-based catalog of 
Herschel-detected sources ($\ge 5\sigma$ detection in at least one of the Herschel bands at the
position of the prior) matched to the COSMOS2015 photometric catalog
\citep{2016ApJS..224...24L} with a search radius of 1\asec (see D17 for more details).
Then, we cross-correlated the COSMOS2015 and i-band catalogs with the radio catalogs
derived from the 0\farcs 9, 1\farcs 2, and 1\farcs 5 resolution-convolved images, finding 
457 new radio sources (not included by D17) that are detected in at least one of the convolved radio
images and with an OIR counterpart within 0\farcs 8. We did not use the 1\farcs8 and 2\farcs2 resolution-convolved images as the small number of sources ($\sim 10$) that could be added from these images
could be significantly affected by spurious identifications.
All these 906 sources that are not included in the 3 GHz radio source counterpart catalog \citep{2017A&A...602A...2S}
based on the full-resolution radio mosaic \citep{2017A&A...602A...1S} were
added to the MR catalog by applying the same method as discussed above to derive
for each one its best resolution and the corresponding total flux density.

\subsection{A multi-resolution catalog}
\label{mr3}
Following the steps outlined in the previous sections, we obtained a catalog of 9\,770 
radio sources with OIR counterparts.
From this catalog we selected only those sources that fulfilled the following criteria.
First, we were interested in investigating the evolution of the linear sizes of different
populations of radio sources, therefore we selected only those with a measured redshift (photometric or
spectroscopic) and for which it was possible to derive a classification  as star-forming galaxy
(SFG) or AGN based on their multiwavelength (X-ray to FIR) properties.
The criteria used to separate these classes are detailed
in \citet{2017A&A...602A...3D} and \citet{2017A&A...602A...2S}.
AGN were further divided into radio-excess (RX) or non-radio-excess (NRX) sources.
Sources showing a radio-excess are defined as those whose radio emission
exceeds by more than $3\sigma$ at any given redshift the star formation rate derived from the
IR through spectral energy distribution (SED) fitting \citep[see][and D17]{2017A&A...602A...1S}.
Based on the multi-wavelength analysis and SED-fitting reported in \citet{2017A&A...602A...3D, 2017A&A...602A...2S},
about 90\% (30\%) of NRX-AGN (RX-AGN)  are moderate-to-high luminosity AGN, so-called HLAGN, and the remaining 10\% (70\%) are
low-to-moderate luminosity AGN, also known as MLAGN. With a reasonable degree of approximation, HLAGN are radiatively efficient AGN
and MLAGN are radiatively inefficient AGN.
Here, by definition, SFGs are always NRX sources (i.e., are only those classified as clean SFG in \citealt{2017A&A...602A...2S}).

\begin{table}
\caption{Number of radio sources in each class in the final MR catalog}
\centering
\begin{tabular}{lrrr}
\hline\hline
  Class     & N    & N$_{\rm res}$ & N$_{\rm unres}$\\
\hline
SFG         & 3\,581 & 2\,765  & 816 \\
NRX-AGN      & 1\,165 &  771  & 394 \\
RX-AGN       & 1\,653 & 1\,062& 591 \\
\hline
\end{tabular}
\tablefoot{Col.1: Classification. Col.2: Total number of sources. Col.3: Number of resolved sources. 
Col.4: Number of unresolved sources}
\label{tab_2}
\end{table}

Second, we restricted the following analysis to sources with a total flux density at 
3\, GHz of $S_{\rm 3GHz}> 20$ $\mu$Jy.
The reason for this choice was that $\sim 90\%$ of the sources with  $S_{\rm 3GHz}<20$ $\mu$Jy
are classified as unresolved in the original catalog because of their low S/N,
compared to $\sim 35\%$ of the sources with $S_T>20$ $\mu$Jy.
For all sources with a best resolution of 0\farcs 75 in the MR catalog, we maintained the classification
into resolved or unresolved that we derived for the full-resolution catalog.
All the other sources, those originally detected at full resolution but 
with a best resolution corresponding to a convolved image and those found only in the
convolved images, were classified as resolved, regardless of the original classification.

With these criteria, we obtained an MR catalog listing 6\,340 single-component and 59
multi-component radio sources. Each source in the MR catalog has an associated best resolution and
a total flux density measured in the best-resolution image.
The number of radio sources divided into different classes is listed in Table\,\ref{tab_2}.
The redshift and 3\,GHz total flux density distributions for the whole sample of 6\,399
sources are shown in Fig.\,\ref{fig:dist1}.
The overall median 3\,GHz total flux density and redshift are
$<S_T>=37$ $\mu$Jy and $<z>=1.0$, respectively.

\begin{figure}[htp]
 \centering
 \includegraphics[width=6.5cm]{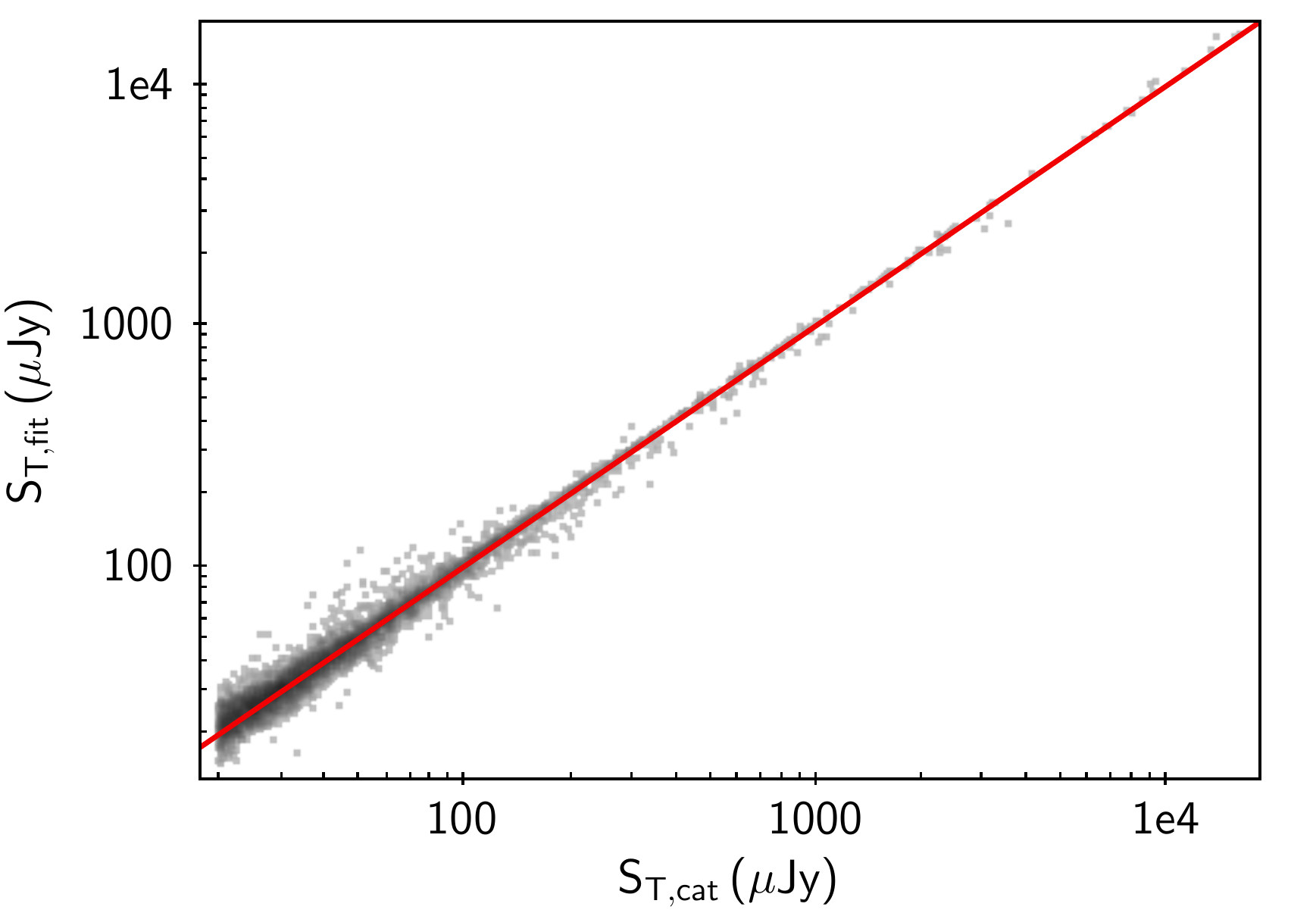}
\caption{Total flux density from a Gaussian fit vs. total flux density from the  MR catalog for all 
the 6\,340 single-component sources. The solid line is the equal flux line.}
 \label{fig:fit}
\end{figure}

\section{Deriving the angular sizes}
\label{A2}

To derive the angular size of each single-component source, we fit a two-dimensional Gaussian component on
its best-resolution image.
We started fitting all the sources  by setting the peak flux to the MR catalog value and not allowing it to vary.
To test the reliability of the fit, we compared the total flux from the fit with that from
the MR catalog.
All  sources whose total flux density from the fit was different by more than 25\% from the value in the
MR catalog were fit again, allowing the peak flux to vary.
For about 90\% of the fitted sources, the flux density from the fit agreed with that
derived from {\tt blobcat}, while the remaining 10\% of the sources had a total flux derived from the
Gaussian fit that was systematically lower than the flux reported in the catalog.
Typically, these sources are heavily resolved and/or strongly asymmetric.
In these cases, {\tt blobcat} is more efficient in  recovering the total
flux because it is based on the flood fill, or thresholding, alghorithm
\citep[][and references therein]{2012MNRAS.425..979H}, which isolates blobs by selecting all pixels above an S/N threshold, without any
a priori assumption on the surface brightness distribution. On the other hand,
Gaussian fits produce more reliable results for slightly resolved sources,
as their surface brightness distribution can be well approximated by a Gaussian.
We solved this discrepancy by fitting these sources
on a lower resolution image and obtained more reliable fits and a much better agreement between
the fitted and cataloged total flux densities.
A small number of fits ($<1\%$) produced clearly incorrect results, and they had to be fit
manually with a more subtle selection of the starting parameters.

Figure\,\ref{fig:fit} shows the comparison between the total flux 
from the MR catalog ($S_{\rm T,cat}$) and the total flux from the one-component Gaussian fit
($S_{\rm T,fit}$) for the 6\,340 single-component sources,.
Defining $R=S_{\rm T,cat}/S_{\rm T,fit}$ as the ratio between the two flux densities, the
median value of R is 1.00, with a median absolute deviation (MAD) of 0.06.
Using a Gaussian fit, with the procedure described above, we were able to reproduce with
excellent agreement
the total flux densities in the catalog that are derived using the flood-fill alghorithm.
From the Gaussian fits we obtained the angular sizes for each source. For the sources
classified as resolved, we used as size measurement the fitted FWHM of the major
axis  deconvolved by the  beam of the image we used for the fit. For the unresolved sources,  
the upper envelope curve in Fig.\,9 from \citet{2017A&A...602A...1S} gives
the upper limit of the total-to-peak flux densities as a function of the S/N.
As discussed in \citet{2017A&A...602A...1S}, the ratio between total and peak flux densities can be used to
estimate the angular size of a radio source in two limiting approximations: a source with
a circular geometry (major and minor axis are equal), or a source with a highly elongated geometry 
in just one direction (minor axis equals zero). We set the upper limit of the sizes of sources
that were classified as unresolved to the geometric mean of the two limiting cases.
Finally, for the 59 sources classified as multiple, we estimated the angular size by measuring the 
source extension directly in the 1\farcs50 resolution mosaic, which has been found to be the most appropriate for
very extended sources.

\section{Converting from Gaussian FWHM into circularized effective radius}
\label{A3}
Radial profiles in SFGs are often fit by an exponential disk or Sersic profile. For a proper comparison, we therefore converted our linear or angular FWHM into the equivalent radius, $R_{e,major}$,
which is the radius containing half of the emitted light along the semimajor axis, using the relation
FWHM$\simeq 2.43 R_{e,major}$ \citep{2017ApJ...839...35M}. We note that this relation is formally valid
for an exponential disk profile (Sersic index $n=1$), while the derived fit values of the Sersic index can have
a large spread and even the median values are typically $>1$.
Furthermore, the data we wish to compare with use the circularized equivalent 
radius (hereafter referred to as R$_e$) to measure the size. The relation to convert  
R$_{e,major}$ into R$_e$ is $R_e=R_{\rm e,major}\sqrt{q}$, where $q$ is
the axial ratio. For our sources we proceeded in the following way. First we derived an average axial 
ratio, $\bar{q}$, for the resolved sources (using a $2\sigma$ limit for the minor axis when undetermined by the fit).
Then, the circularized R$_e$ for the resolved sources was calculated as $R_e=FWHM\times \sqrt{\bar{q}}/2.43$,
while for the upper limits, $R_e=FWHM/2.43$.  

\section{Tables}
\begin{table}
\caption{Median FWHM radio size redshift evolution of $\mu$Jy populations}
\label{tab-a1}
\centering
\begin{tabular}{lcccrrr}
  \hline\hline
  Class  &$z_{\rm min}$&$z_{\rm max}$&$z_{\rm med}$& N & N$_{\rm unres}$& FWHM    \\
         &           &           &           &   &             &  (kpc) \\
  \hline
  SFG    & 0.00      & 0.20      &  0.14     &  163 &   8       & $2.82^{+0.38}_{-0.22}$ \\
  SFG    & 0.20      & 0.35      &  0.27     &  340 &  24       & $3.96^{+0.41}_{-0.34}$ \\
  SFG    & 0.35      & 0.55      &  0.44     &  480 &  46       & $4.54^{+0.25}_{-0.20}$ \\
  SFG    & 0.55      & 0.75      &  0.68     &  465 &  82       & $4.96^{+0.21}_{-0.21}$ \\
  SFG    & 0.75      & 1.00      &  0.89     &  568 & 114       & $4.93^{+0.20}_{-0.16}$ \\
  SFG    & 1.00      & 1.40      &  1.20     &  543 & 148       & $4.80^{+0.12}_{-0.20}$ \\
  SFG    & 1.40      & 1.80      &  1.57     &  366 & 112       & $4.43^{+0.20}_{-0.24}$ \\
  SFG    & 1.80      & 2.50      &  2.03     &  410 & 128       & $4.27^{+0.27}_{-0.23}$ \\
  SFG    & 2.50      & 7.00      &  2.93     &  246 &  97       & $3.57^{+0.20}_{-0.28}$ \\
  \hline
  NRX-AGN& 0.00      & 0.30      &  0.22     &   42 &  11       & $1.59^{+0.2}_{-0.6}$ \\
  NRX-AGN& 0.30      & 0.60      &  0.42     &  159 &  46       & $2.45^{+0.28}_{-0.15}$ \\
  NRX-AGN& 0.60      & 0.90      &  0.75     &  188 &  61       & $3.51^{+0.37}_{-0.30}$ \\
  NRX-AGN& 0.90      & 1.20      &  1.04     &  200 &  70       & $3.14^{+0.16}_{-0.18}$ \\
  NRX-AGN& 1.20      & 1.50      &  1.35     &  154 &  65       & $3.11^{+0.40}_{-0.13}$ \\
  NRX-AGN& 1.50      & 2.00      &  1.70     &  150 &  48       & $3.49^{+0.80}_{-0.15}$ \\
  NRX-AGN& 2.00      & 7.00      &  2.58     &  272 &  93       & $3.57^{+0.17}_{-0.13}$ \\
  \hline
  RX-AGN & 0.00      & 0.30      &  0.22     &   53 &  13       & $1.12^{+0.35}_{-0.44}$ \\
  RX-AGN & 0.30      & 0.60      &  0.43     &  183 &  64       & $1.47^{+0.11}_{-0.15}$ \\
  RX-AGN & 0.60      & 0.90      &  0.74     &  339 & 130       & $1.74^{+0.12}_{-0.10}$ \\
  RX-AGN & 0.90      & 1.20      &  1.02     &  275 &  93       & $2.06^{+0.20}_{-0.30}$ \\
  RX-AGN & 1.20      & 1.50      &  1.35     &  205 &  77       & $2.43^{+0.37}_{-0.20}$ \\
  RX-AGN & 1.50      & 2.00      &  1.73     &  294 & 117       & $2.49^{+0.16}_{-0.23}$ \\
  RX-AGN & 2.00      & 7.00      &  2.56     &  304 &  97       & $2.79^{+0.23}_{-0.18}$ \\
\end{tabular}
\tablefoot{Col.1: Classification. Cols.2 and 3: Redshift bin. Col.4: Median redshift. Col.5: Total number of sources. Col.6:
  Number of unresolved sources (upper limits). Col.7: Median FWHM linear size in Kpc with errors.}
\end{table}

\begin{table}
\caption{Median FWHM radio size  luminosity evolution of $\mu$Jy populations}
\label{tab-a2}
\centering
\begin{tabular}{lcccrrr}
  \hline\hline
  Class  &$\log{L_{\rm 3GHz,min}}$&$\log{L_{\rm 3GHz,max}}$&$\log{L_{\rm 3GHz,med}}$& N & N$_{\rm unres}$& FWHM    \\
         & (W/Hz)              &  (W/Hz)             &  (W/Hz)             &   &              & (kpc)  \\
  \hline
  SFG    & 18.00      & 22.00      &  21.8     &  292 &  31       & $2.90^{+0.20}_{-0.10}$ \\
  SFG    & 22.00      & 22.50      &  22.3     &  511 &  45       & $4.71^{+0.25}_{-0.20}$ \\
  SFG    & 22.50      & 23.00      &  22.8     &  855 & 156       & $4.93^{+0.17}_{-0.10}$ \\
  SFG    & 23.00      & 23.50      &  23.2     &  873 & 214       & $4.96^{+0.24}_{-0.15}$ \\
  SFG    & 23.50      & 23.75      &  23.6     &  402 & 119       & $4.54^{+0.26}_{-0.21}$ \\
  SFG    & 23.75      & 24.00      &  23.9     &  331 & 100       & $4.05^{+0.13}_{-0.15}$ \\
  SFG    & 24.00      & 24.50      &  24.2     &  294 &  88       & $3.54^{+0.26}_{-0.20}$ \\
  SFG    & 24.50      & 27.00      &  24.6     &   23 &   5       & $3.57^{+1.00}_{-0.60}$ \\
  \hline
  NRX-AGN& 18.00      & 22.50      &  22.2     &  137 &  47       & $2.22^{+0.30}_{-0.17}$ \\
  NRX-AGN& 22.50      & 23.00      &  22.8     &  200 &  74       & $3.08^{+0.38}_{-0.15}$ \\
  NRX-AGN& 23.00      & 23.50      &  23.3     &  305 & 113       & $3.35^{+0.30}_{-0.18}$ \\
  NRX-AGN& 23.50      & 23.75      &  23.6     &  161 &  52       & $3.84^{+0.30}_{-0.30}$ \\
  NRX-AGN& 23.75      & 24.00      &  23.9     &  142 &  51       & $3.15^{+0.16}_{-0.20}$ \\
  NRX-AGN& 24.00      & 24.50      &  24.2     &  187 &  54       & $3.46^{+0.18}_{-0.34}$ \\
  NRX-AGN& 24.50      & 27.00      &  24.7     &   33 &   3       & $2.57^{+0.90}_{-0.80}$ \\
  \hline
  RX-AGN & 18.00      & 22.50      &  22.3     &  104 &  45       & $1.35^{+0.20}_{-0.45}$ \\
  RX-AGN & 22.50      & 23.00      &  22.8     &  212 & 105       & $1.47^{+0.30}_{-0.15}$ \\
  RX-AGN & 23.00      & 23.50      &  23.3     &  340 & 163       & $1.86^{+0.26}_{-0.10}$ \\
  RX-AGN & 23.50      & 23.75      &  23.6     &  268 & 115       & $2.16^{+0.27}_{-0.18}$ \\
  RX-AGN & 23.75      & 24.00      &  24.9     &  232 &  72       & $2.27^{+0.48}_{-0.21}$ \\
  RX-AGN & 24.00      & 24.50      &  24.2     &  300 &  75       & $2.51^{+0.10}_{-0.16}$ \\
  RX-AGN & 24.50      & 27.00      &  24.9     &  197 &  16       & $2.18^{+0.24}_{-0.19}$ \\
\end{tabular}
\tablefoot{Col.1: Classification. Cols.2 and 3: 3\,GHz radio luminosity log bin. Col.4: Median 3\,GHz radio luminosityt.
  Col.5: Total number of sources. Col.6:
  Number of unresolved sources (upper limits). Col.7: Median FWHM iinear size in Kpc with errors.}
\end{table}
\end{appendix}
\end{document}